\documentclass[%
 aip,
 amsmath,amssymb,
 reprint,%
]{revtex4-1}

\usepackage{graphicx}
\usepackage{dcolumn}
\usepackage{bm}
\usepackage{tikz}
\usepackage{tikz-feynman}
\usepackage{xcolor,slashed}
\usepackage[utf8]{inputenc}
\usepackage[T1]{fontenc}
\usepackage{mathptmx}
\usepackage{etoolbox}
\usepackage{adjustbox}
\usepackage{hyperref}
\usepackage{cleveref}
\DeclareUnicodeCharacter{2212}{-}

\makeatletter
\def\@email#1#2{%
 \endgroup
 \patchcmd{\titleblock@produce}
  {\frontmatter@RRAPformat}
  {\frontmatter@RRAPformat{\produce@RRAP{*#1\href{mailto:#2}{#2}}}\frontmatter@RRAPformat}
  {}{}
}%
\makeatother
\begin{document}

\preprint{AIP/123-QED}

\title{Enhancing Sensitivity in Ge-Based Rare-Event Physics Experiments through Underground Crystal Growth and Detector Fabrication}
\author{Dongming Mei}
 \email{dongming.mei@usd.edu}
\affiliation{ 
Physics Department, University of South Dakota, Vermillion, SD, 57069}

\date{\today}

\begin{abstract}

The cosmogenic production of long-lived isotopes such as \(^{3}\)H,\(^{55}\)Fe, \(^{60}\)Co, \(^{65}\)Zn, and \(^{68}\)Ge poses a significant challenge as a source of background events in Ge-based dark matter (DM) and neutrinoless double-beta decay (\(0\nu\beta\beta\)) experiments. In the pursuit of DM, particularly within the largely unexplored parameter space for low-mass DM, new detector technologies are being developed with extremely low-energy thresholds to detect MeV-scale DM. However, isotopes like \(^{3}\)H, \(^{55}\)Fe, \(^{65}\)Zn, and \(^{68}\)Ge, produced cosmogenically within the detector material, emerge as dominant backgrounds that severely limit sensitivity in these searches. Similarly, efforts to detect \(0\nu\beta\beta\), especially under a neutrino normal mass hierarchy scenario, require a sensitivity to the effective Majorana mass of $\sim$1 meV. Achieving this level of sensitivity necessitates stringent suppression of background signals from isotopes such as \(^{60}\)Co and \(^{68}\)Ge, which impose critical detection limits. To reach the targeted sensitivity for these next-generation experiments and to unlock their full discovery potential for both low-mass DM and \(0\nu\beta\beta\), relocating Ge crystal growth and detector fabrication to underground environments is crucial. This approach is the most effective strategy to significantly reduce the production of these long-lived isotopes, thereby ensuring the experimental sensitivity required for groundbreaking discoveries.

\end{abstract}

\maketitle

\section{Introduction}

The search for dark matter (DM) and the quest to observe neutrinoless double-beta decay (\(0\nu\beta\beta\)) represent two of the most significant challenges in modern physics. Germanium (Ge)-based detectors have emerged as leading tools in these searches due to their excellent energy resolution and the ability to achieve low background levels \cite{supercdms, cdex, edelweiss, gerda, majorana, agostini_review_2019}. However, one of the critical limitations to the sensitivity of these detectors is the production of cosmogenic isotopes, such as tritium (\(^{3}\)H), iron-55 (\(^{55}\)Fe), cobalt-60 (\(^{60}\)Co), zinc-65 (\(^{65}\)Zn), and Ge-68 (\(^{68}\)Ge), which are produced when Ge detectors are exposed to cosmic rays during the crystal growth and detector fabrication processes at the Earth's surface\cite{mei_cosmogenic_2009, wei2017, cebrian2017}.

Cosmogenic isotopes present a significant source of background in Ge-based DM experiments, particularly in the search for low-mass DM candidates, where a vast unexplored parameter space exists for masses in the MeV range. These isotopes can mimic the signal of DM interactions by producing low-energy events, thereby setting a fundamental limit on the sensitivity of the detectors \cite{cebrian2017}. This challenge also arises in \(0\nu\beta\beta\) experiments, where the presence of cosmogenic isotopes such as \(^{60}\)Co and \(^{68}\)Ge can produce background events that interfere with the detection of the extremely rare \(0\nu\beta\beta\) decay \cite{abgrall_majorana_2014}.

To achieve the target sensitivity required for these experiments, it is crucial to mitigate the production of cosmogenic isotopes. State-of-the-art Ge-based experiments, such as SuperCDMS~\cite{supercdms} for DM searches and LEGEND~\cite{legend} for 0$\nu\beta\beta$ decay, must implement effective strategies to address cosmogenic backgrounds. For instance, SuperCDMS~\cite{supercdms} not only requires the rapid transportation of detectors from the Earth's surface to minimize exposure but also depends on its exceptional ability to distinguish between electronic recoils (e-recoils) and nuclear recoils (n-recoils). This capability significantly reduces background noise from cosmogenically produced events, thereby enhancing the search for DM with masses greater than $\sim$1 GeV/c$^2$. However, in the pursuit of low-mass DM using High Voltage detectors, this e/n recoil discrimination capability diminishes, resulting cosmogenic backgrounds to dominate the sensitivity. Conversely, LEGEND~\cite{legend} imposes stringent surface exposure limits, restricting exposure to just 30 days. This approach effectively limits the contribution of cosmogenic backgrounds to only 20\% of the total background budget, thereby preserving the experiment's sensitivity to a Majorana effective mass of $\sim$10 meV. However, for future experiments beyond LEGEND-1000, which aim to probe the Majorana effective mass down to approximately 1 meV, it will be necessary to reduce this cosmogenic background by an additional factor of 30~\cite{mei2024}.

If not the only practical solution, the most effective strategy for achieving the required sensitivity in next-generation Ge-based experiments is to relocate Ge crystal growth and detector fabrication to underground environments. Underground facilities are naturally shielded from cosmic rays by the Earth's crust, significantly reducing the rate of cosmogenic activation \cite{hehn_cosmogenic_2014}. By minimizing the exposure of Ge detectors to cosmic radiation, it is possible to suppress the production of long-lived isotopes, thereby enhancing the sensitivity of the detectors to both DM interactions and 
0$\nu\beta\beta$ decay events \cite{aalseth_search_2018}.

This paper examines the impact of cosmogenic isotope production on the sensitivity of next-generation Ge-based experiments and underscores the necessity and feasibility of creating underground environments for Ge crystal growth and detector fabrication to ensure the success of these critical searches.

Currently, cosmogenic background constitutes only a small fraction of the total background in existing Ge-based experiments like SuperCDMS and LEGEND. However, as other backgrounds are mitigated, cosmogenic background is likely to become a dominant source in future experiments. This paper assumes that within the next 10 years, substantial progress will be made in controlling these other background sources. For instance, experimental sensitivity has improved by one to two orders of magnitude each decade, driven by advancements in detector technologies, material production and purification, and software innovations in data analysis. We anticipate similar advancements over the next decade, potentially making cosmogenic production on the surface the limiting factor.

While achieving this is challenging, it is essential for detecting low-mass dark matter at the MeV scale and progressing beyond the current ton-scale in 0$\nu\beta\beta$ decay experiments. Given that developing the necessary technology takes about a decade, it is crucial to begin constructing the underground Ge crystal growth and detector fabrication facility now.

\section{Impact of Cosmogenic Isotope Production on the Sensitivity of Next-Generation Ge-Based Experiments}

\begin{figure}[h]
    \centering
    \includegraphics[width=0.55\textwidth]{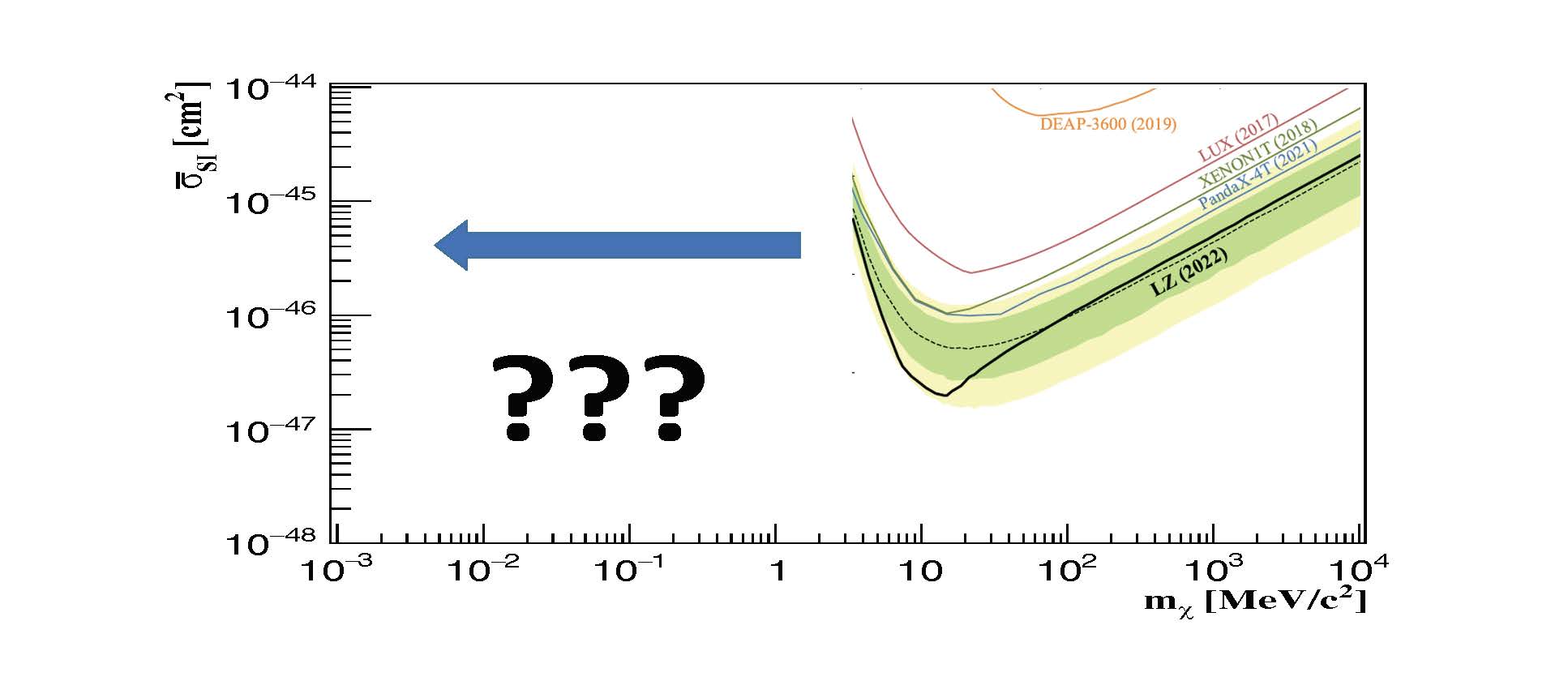}
    \caption{DM-nucleon cross section as a function of DM mass, revealing a vast unexplored parameter space that underscores the need for more sensitive experiments. }
    \label{fig:fig1}
\end{figure}

Weakly Interacting Massive Particles (WIMPs) \cite{smith_wimps_1990} have long been considered a leading candidate for DM. These particles, with masses thought to be comparable to those of heavy nuclei, interact with atomic nuclei via extremely weak and short-range forces. Although WIMPs are expected to collide with atomic nuclei only very rarely, such collisions would impart significant recoil energy to the nuclei, causing them to recoil at velocities several thousand times the speed of sound \cite{lewin_wimp_nucleus_1996}. Numerous experiments have been designed to directly detect the small recoil energies resulting from WIMP-nucleus interactions \cite{cdex, edelweiss, agnese_supercdms_2018, aprile_xenon1t_2018, agnese_supercdms_2019, abramoff_darkside_2020, akerib_lux_2017, ahmed_cdms_2010, deap, pandax}. The LZ experiment \cite{akerib_lz_2020}, currently collecting data at the Sanford Underground Research Facility (SURF), has provided the best experimental sensitivity to date, ruling out a significant portion of the parameter space previously allowed for WIMP detection (as shown in Figure~\ref{fig:fig1}).

In the past decade, light, MeV-scale DM \cite{essig_dark_2012, lee_thermally_2015, finkbeiner_mev_dm_2016} has emerged as an intriguing alternative DM candidate. Despite considerable recent efforts in searching for DM-electron interactions \cite{agnes_darkside_2018, aguilar_fermi_2020, abdelhameed_cresst_2019, armengaud_eres_2020, abrams_dmtpc_2019, agnes_darkside_2021, aramaki_prospects_2016, an_aboveground_2017, cao_darkside_2021}, a large portion of the parameter space for MeV-scale DM remains unexplored~\cite{cao_darkside_2021}, as shown in Figure~\ref{fig:fig1}. Detecting MeV-scale DM requires novel detectors with extremely low-energy thresholds (below 100 eV) and minimal background noise. Ge-based experiments are poised to play a crucial role in the search for sub-GeV and MeV-scale DM.

To demonstrate the existence of Majorana neutrinos, numerous \(0\nu\beta\beta\) experiments have been conducted over the past several decades \cite{experiment1, experiment2, experiment3, experiment4, KamLandZen, gerda_2018}. Despite more than 30 years of dedicated research, \(0\nu\beta\beta\) decay has yet to be observed. The most stringent constraint on the decay half-life, currently at $\sim$10$^{26}$ years, has been set by the KamLANDZen~\cite{KamLandZen} and GERDA experiments, where GERDA is a Ge-based project \cite{gerda_2018}. Looking forward, the planned ton-scale Ge-based experiment, LEGEND-1000, aims to achieve a sensitivity that surpasses \(10^{28}\) years for the \(0\nu\beta\beta\) decay half-life \cite{legend_2020_1}.

Cosmogenic production of long-lived radioactive isotopes in Ge during crystal growth and detector fabrication on the Earth's surface has a significant impact on the sensitivity of next-generation Ge-based DM and \(0\nu\beta\beta\) experiments. These isotopes, produced through interactions with cosmic rays, set stringent limits on the discovery potential of these experiments, as observed in several major experimental efforts, including SuperCDMS, the Majorana Demonstrator, and GERDA \cite{supercdms_2014, majorana_2015, gerda_2017}.

For instance, the SuperCDMS experiment has detected cosmogenically produced isotopes such as \(^{3}\)H, \(^{55}\)Fe, \(^{65}\)Zn, and \(^{68}\)Ge in their detectors, with production rates measured at 74$\pm$9, 1.5$\pm$0.7, 17$\pm$5, and 30$\pm$18 atoms/kg/day, respectively \cite{supercdms_isotopes_2016}. These findings show reasonable alignment with Monte Carlo simulations, particularly those conducted by Wei et al. \cite{wei2017}, which utilized cross-section data from ACTIVIA 1/2 (as detailed in their Table 3). The simulations predict that these isotopes are primarily produced through neutron interactions with Ge isotopes, using a neutron energy spectrum representative of sea-level conditions. The observed production rates are within a factor of 2 to 3 of the simulated values (34.12/52.37, 3.29/4.10, 19.53/44.19, and 10.25/24.65 for $^{3}$H, $^{55}$Fe, $^{65}$Zn, and $^{68}$Ge, respectively, based on ACTIVIA 1/2 cross-section data), which is encouraging given the significant uncertainties in the interaction cross sections.

The production of cosmogenic isotopes, particularly in surface-based facilities, remains a dominant source of background events in DM searches, potentially overwhelming the rare signals that experiments like SuperCDMS seek to detect low-mass DM. Without mitigation, trace-level production of radioisotopes during surface processing of Ge and Si crystals could severely compromise the sensitivity of future DM experiments \cite{supercdms, mei2018}.

\begin{figure}[h]
    \centering
    \includegraphics[width=0.55\textwidth]{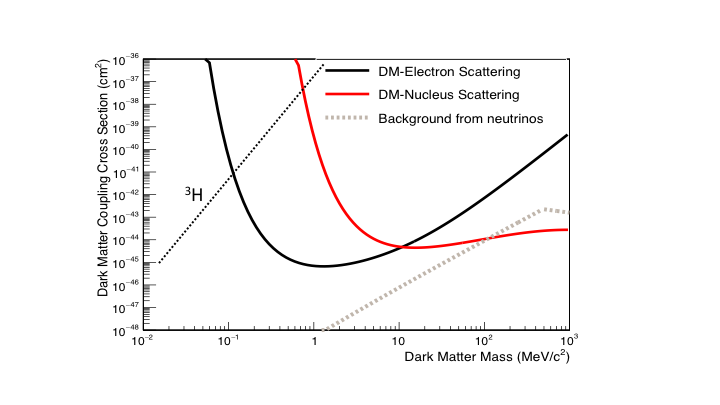}
    \caption{The projected experimental sensitivity for one-kg year is shown, assuming that background events are limited solely by $^3$H and neutrino-induced interactions. }
    \label{fig:fig3}
\end{figure}

Figure~\ref{fig:fig3} illustrates the sensitivity of low-mass DM searches using a hypothetical highly sensitive Ge-based detector with internal charge amplification, achieving an exceptionally low-energy threshold down to 0.1 eV~\cite{mei2018}. The formulas used to generate this figure are detailed in Mei et al.~\cite{mei2018}. As depicted in the figure, background events produced by $^{3}$H significantly limit the sensitivity, even before accounting for contributions from shorter-lived isotopes such as $^{56}$Fe, $^{65}$Zn, and $^{68}$Ge. This highlights the critical need to mitigate cosmogenic production in order to effectively explore low-mass DM.

\begin{figure}[h]
    \centering
    \includegraphics[width=0.45\textwidth]{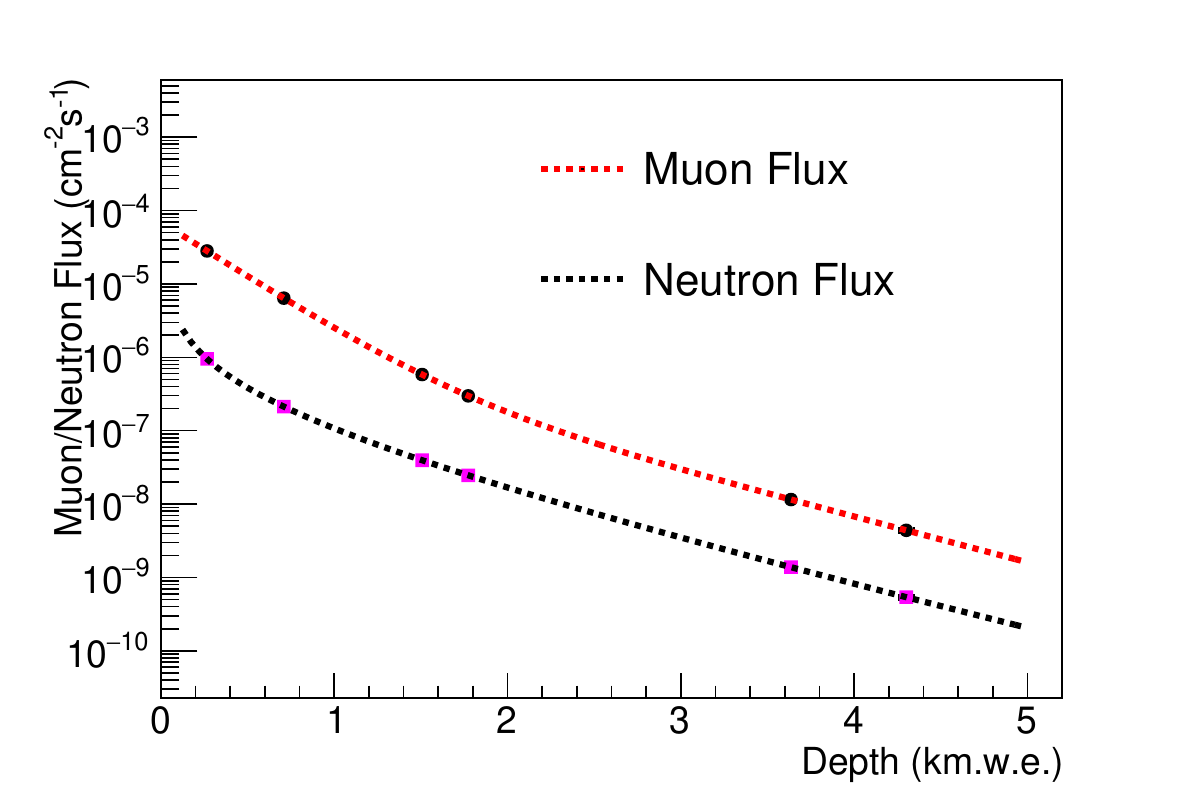}
    \caption{Muon and neutron flux at different levels at SURF. The black dots on the red dashed line are the predicted muon-flux at the depth of 300-ft (0.266 km.w.e.), 800-ft (0.709 km.w.e.), 1700-ft (1.507 km.w.e.), 2000-ft (1.773 km.w.e.), 4100-ft (3.635 km.w.e.), and 4850-ft (4.3 km.w.e), respectively using the mode described in Mei-Hime paper~\cite{mei}. The magenta triangles on the black dashed line represent the predicted neutron flux at the corresponding levels at SURF. }
    \label{fig:fig2}
\end{figure}

The most effective strategy to mitigate cosmogenic isotope production is to relocate Ge crystal growth and detector fabrication to underground facilities. This approach significantly reduces exposure to cosmic rays, particularly muon-induced neutron fluxes, which are reduced by more than five orders of magnitude at the 4850-ft level (4.3 km.w.e.) at SURF \cite{mei}. Figure \ref{fig:fig2} illustrates the substantial reduction in muon-induced neutron flux with depth, demonstrating the potential for underground environments to effectively minimize cosmogenic activation.

Moreover, the projected background from isotopes like \(^{60}\)Co and \(^{68}\)Ge constitutes about 20\% of the LEGEND-1000 background budget, highlighting the critical need for underground processing to achieve the necessary sensitivity for future large-scale experiments \cite{legend_2020}. By performing Ge purification, crystal growth, characterization, and detector fabrication entirely within underground laboratories at the 4850-ft level at SURF, the production of cosmogenically induced isotopes can be rendered negligible, thereby allowing the next generation experiments to reach unprecedented levels of sensitivity. 

\begin{figure}[h]
    \centering
    \includegraphics[width=0.5\textwidth]{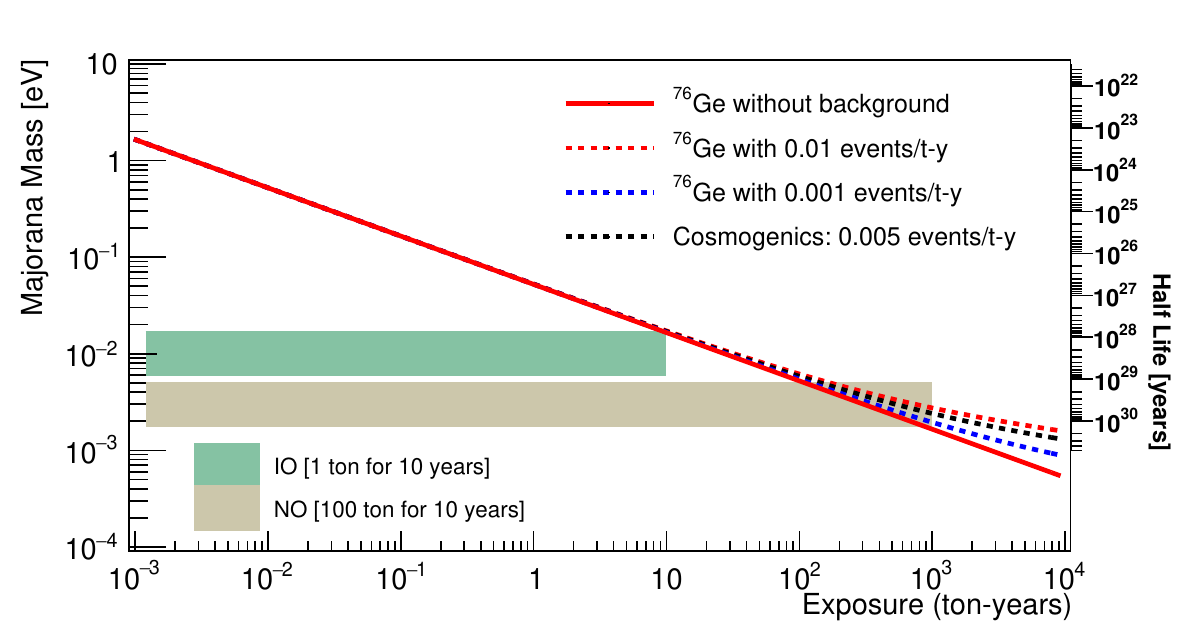}
    \caption{The figure displays the projected sensitivity for a prospective 100-ton scale $^{76}$Ge experiment, utilizing technology similar to that of LEGEND-1000. The dimensions of the shaded regions represent the assumed exposure and the parameter space for the minimum neutrino mass ($m_{L}$), under the assumption that the decay is mediated by a light Majorana neutrino.}
    \label{fig:fig4}
\end{figure}

Figure~\ref{fig:fig4} presents the projected sensitivity of Ge-based experiments in exploring both normal and inverted mass hierarchies of neutrinos~\cite{mei2024}, taking into account cosmogenic production on the surface. The equations used to derive this figure are detailed in Mei et al.\cite{mei2024}. As shown in the figure, the 20\% background contribution from surface cosmogenic production is not a significant issue for a ton-scale experiment like LEGEND-1000. However, it does limit the sensitivity of a 100-scale experiment aimed at achieving a $\sim$1 meV Majorana effective mass in the normal mass hierarchy. This indicates that a factor of $\sim$30 reduction in cosmogenic background is necessary to have discovery potential for 0$\nu\beta\beta$ decay experiments if the decay is mediated by the minimum neutrino mass~\cite{mei2024}.

The significance of reducing various backgrounds was highlighted in the 2022 Snowmass reports on direct dark matter searches and $0\nu\beta\beta$ decay \cite{snowmass_2022}. These reports stress the need for next-generation experiments and corresponding facilities to enhance sensitivity for both dark matter detection and 0$\nu\beta\beta$ decay. An underground crystal growth facility would be ideally suited to meet these crucial requirements. The Snowmass recommendations underscore the necessity of moving key stages of detector development underground to fully unlock the discovery potential of next-generation experiments.

\section{The Feasibility of Underground Ge Crystal Growth and Detector Fabrication}

As noted by Wei et al.\cite{wei2017}, cosmogenic isotopes are primarily generated by neutrons. At a depth of 800 feet, equivalent to 0.709 km.w.e., the flux of these neutrons is reduced by over four orders of magnitude compared to the Earth's surface, effectively minimizing cosmogenic production at this depth (see Figure\ref{fig:fig2}). SURF provides a range of depth levels, making it an ideal location for establishing an advanced facility where Ge purification, crystal growth, crystal characterization, and detector fabrication can all be conducted in underground laboratories at the 800-ft level. This environment is crucial for meeting the stringent requirements of next-generation experiments.

The long-lived isotopes produced at the Earth's surface in Ge ingots will be largely removed during the Ge purification process through zone refining. To fully realize the potential of this environment, it is necessary to establish a comprehensive underground facility that includes Ge purification, crystal growth, mechanical processing and characterization, and detector fabrication.

Ge is a rare element in the Earth’s crust, with an estimated abundance of approximately 7 parts per million (ppm). Ge is primarily produced as a byproduct of zinc ore processing and through the extraction from coal fly ash. Natural germanium comprises five isotopes: $^{70}$Ge (20.52\%), $^{72}$Ge (27.45\%), $^{73}$Ge (7.76\%), $^{74}$Ge (36.52\%), and $^{76}$Ge (7.75\%).

As a semiconductor material, Ge has a wide range of industrial applications, including electronics, fiber optic systems, infrared optics, polymerization catalysts, and solar technologies. However, commercially available Ge ingots typically have impurity levels ranging from 99.99\% to 99.9999\%. Even the highest purity level commercially available is insufficient for growing Ge crystals intended for detector production.

Therefore, commercial ingots must undergo further purification to reach an ultra-high purity level of 99.999,999,999,9\% to grow a Ge crystal. Only then does a portion of this crystal have the potential to achieve the extreme purity level required to fabricate a Ge detector, with an impurity level of 99.999,999,999,99\%.

\begin{figure}[h]
    \centering
    \includegraphics[width=0.45\textwidth]{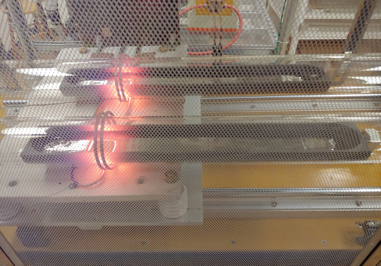}
    \caption{Shown is the Ge purification process through zone refining at USD~\cite{zone1, zone2, zone3}. }
    \label{fig:fig5}
\end{figure}

Ge is purified using a process called zone refining (Figure~\ref{fig:fig5}), which involves creating a melting zone that travels along the length of the Ge ingot. This process, first developed at Bell Labs in 1954, works by segregating impurities at the boundary between the liquid and solid phases within the melting zone. As the melting zone moves, impurities remain in the liquid phase and are transported to the end of the ingot. After multiple passes,  the impurities become concentrated at the far end of the ingot, which is then removed to increase the purity of the remaining Ge.

Through meticulous quality control and the optimization of various parameters, the zone refining process can purify Ge ingots from 99.99\% to an extraordinary 99.999,999,999,9\%, achieving an eight orders of magnitude reduction in impurities. This process effectively removes cosmogenically produced long-lived isotopes from the Earth's surface, making the purified Ge ingots ideal for Ge crystal growth. After 15 years of research and development (R\&D) at the surface lab \cite{zone1, zone2, zone3}, the University of South Dakota (USD) has established a standardized procedure that consistently achieves a high yield (~80\%) of purified Ge ingots that meet the stringent requirements for crystal growth. This R\&D program positions USD as a significant contributor to the global production of qualified Ge crystals, alongside a handful commercial companies. As a research institution, USD is uniquely positioned to transfer this technology to underground Ge production, paving the way for the next generation of Ge-based experiments.

Large-size Ge crystals are grown using the Czochralski technique, which was developed in the 1930s. Crystal growth is a highly intricate process involving heat, momentum, and mass transport phenomena, along with chemical reactions (e.g., contamination of crystal and melt) and electromagnetic processes (e.g., induction and resistance heating, magnetic stirring, and magnetic breaks). Consequently, the crystal growth process is dynamic, involving phase transformations from liquid to solid. The interface control between the liquid and solid phases occurs on the nanometer scale, while the growth system itself spans approximately a meter in size. This complexity requires the optimization of numerous parameters (10 or more), each with its own set of constraints. As a result, the dynamic nature of crystal growth is challenging to control.

The growth rate and quality of high-purity Ge crystals largely depend on the precise control of the thermal field (heat transfer and temperature profile). However, these control parameters can only be regulated externally, through the geometry of the growth system, gas flow rate and pressure, pulling rate, frequency, and the power of the RF heater. Measurements inside the growth chamber, where temperatures exceed 1000 $^{\circ}$C, and the quantitative determination of control parameters are technically challenging.

\begin{figure}[h]
    \centering
    \includegraphics[width=0.45\textwidth]{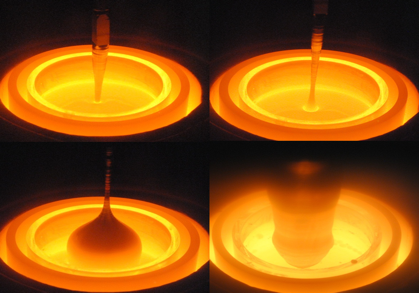}
    \caption{Ge crystal growth stages: Upper left: seeding; Upper right: dashing; Lower left: body growth; Lower right: growing to dry. }
    \label{fig:fig6}
\end{figure}

After 15 years of R\&D at the surface lab \cite{growth1, growth2, growth3, growth4, growth5}, USD has invented a growth method (patent number 10,125,431) that consistently facilitates the production of detector-grade crystals (99.999,999,999,999,9\% purity) on a regular basis (Figure~\ref{fig:fig6}). This advancement paves the way for growing Ge crystals underground.

Since 2009, the USD group has published research \cite{growth1, growth2, growth3, growth4, growth5} demonstrating improvements in the quality of large-size Ge crystals, highlighting our ability to control the parameters necessary for the growth of low-dislocation (3,000–7,000 etch pits/cm$^2$), large-diameter ($\sim$12 cm), and high-purity Ge single crystals ($\sim 10^{10}$/cm$^2$) for detector fabrication. 

After the purification of Ge ingots, three slices are typically cut along the zone-refined ingots to measure the impurity distribution using the Hall Effect method. This method determines both the impurity level and charge mobility. If the impurity level ($<2\times10^{11}$/cm$^3$) and mobility ($>3\times10^4$ cm$^2$/Vs) meet the requirements for crystal growth, that portion of the zone-refined ingot is selected for Ge crystal growth. At USD, the usable portion of the zone-refined ingots for crystal growth usually ranges between 70\% and 80\% of the entire ingot.

Additionally, when Ge crystals are grown, slices are cut from various parts of the crystal, including the neck, shoulder, middle, and end, to measure impurity distribution using a Hall Effect system. If a portion of the grown crystal meets the necessary criteria—(1) an impurity level between 5$\times10^9$/cm$^3$ to 3$\times$10$^{10}$/cm$^3$ , (2) mobility greater than 45,000 cm\(^2\)/Vs, and (3) a dislocation density between 100 and 7,000 etch-pits/cm\(^2\)—this portion is deemed suitable for making a Ge detector. The crystal is then mechanically processed into the desired geometry for making electrical contacts, necessitating a mechanical processing and crystal characterization lab.

\begin{figure}[h]
    \centering
    \includegraphics[width=0.45\textwidth]{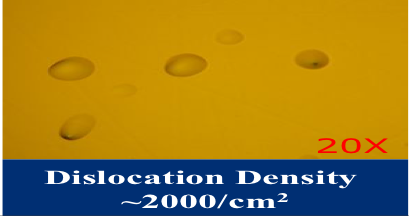}
    \caption{ Measured dislocations from a USD-grown crystal. The dots with circles are dislocations. 
}
    \label{fig:fig7}
\end{figure}

Characterizing grown crystals provides crucial feedback for the growth process and helps determine whether the crystals are of sufficient quality for detector fabrication. With 15 years of R\&D experience, USD has established a comprehensive characterization program to assess whether crystals meet the qualifications for detector fabrication. The quality of a single-crystalline Ge crystal is determined through X-ray diffraction and dislocation density measurements using an optical microscope. Figure~\ref{fig:fig7} shows an example of the dislocation density measured via microscopy for a crystal grown at USD.

Commercially available Ge detectors are traditionally fabricated using lithium diffusion for the n+ contact and boron implantation for the p+ contact. The fundamental principle behind this method is to create a charge barrier through a conventional p-n junction. More recently, scientists at Lawrence Berkeley National Laboratory (LBNL) have developed a bi-polar blocking technology that utilizes amorphous Ge and amorphous Si as detector contacts~\cite{mark}. This innovative technology enables the creation of thin contacts ($\sim$600 nm) on Ge, compared to the much thicker Li-diffused contacts ($\sim$1 mm). One significant advantage of thin contacts is their suitability for fabricating segmented Ge detectors. In collaborating with LBNL, USD has developed the capability to produce thin-contact detectors using sputtering technology. After 7 years of R\&D, USD has established a robust detector fabrication process and successfully demonstrated thin-contact technology with USD-grown crystals \cite{de1, de2, de3, de4, de5, de6, de7}. A detailed description of the fabrication processes used to transform single-crystal Ge or Si boules into functional detectors—including crystal alignment, shaping, polishing, and sensor fabrication—can be found in Ref. \cite{de2}.

\section{Conclusion}
In summary, cosmogenic isotope production is a significant limiting factor for the sensitivity of future Ge-based DM and \(0\nu\beta\beta\) decay experiments. Relocating the critical processes of crystal growth and detector fabrication to underground environments can substantially reduce these backgrounds, thereby enhancing the experimental sensitivity required for groundbreaking discoveries. The pursuit of DM and \(0\nu\beta\beta\) decay detection demands cutting-edge technologies capable of producing the most sensitive detectors, a challenge that requires decades of dedicated research and development.

The United States is well-positioned to advance Ge crystal growth and detector development, building on technology that originated at LBNL. With the guidance and mentorship of LBNL pioneers, USD has successfully developed its own expertise in Ge crystal growth and detector fabrication. Over nearly 15 years of continuous R\&D, supported by the Department of Energy, the National Science Foundation, and the state of South Dakota, USD has refined this technology to the point where the production of high-purity Ge crystals and detectors in an underground environment is now feasible. This progress not only builds on the foundational work at LBNL but also positions the United States at the forefront of next-generation experiments in DM detection and 0$\nu\beta\beta$ decay, where ultra-pure Ge crystals and highly sensitive detectors are crucial.

Additionally, institutions within the SuperCDMS collaboration, such as SLAC and Texas A\&M, have developed their own detector fabrication capabilities, further bolstering the United States' leading position in producing cutting-edge detectors for frontier science. Complementing these efforts, scientists at South Dakota Mines have pioneered advanced radon reduction techniques to remove radon from the air, ensuring that detectors fabricated underground have radon daughter plate-out free contacts. Together, these advancements position the United States to establish a state-of-the-art underground facility for Ge crystal growth and detector development, meeting the stringent sensitivity requirements of next-generation experiments.

SURF in South Dakota is the only underground lab of its kind in the United States, offering unique access to multiple depths, including 4,850 feet below the surface. This versatility in depth makes SURF an ideal location for establishing an underground Ge crystal growth and detector fabrication facility. The deep underground environment provides exceptional shielding from cosmic rays, significantly reducing the production of cosmogenic isotopes and ensuring the high purity Ge crystals and detectors required for next-generation Ge-based experiments in DM detection and 0$\nu\beta\beta$ decay.

In conclusion, the establishment of an underground crystal growth and detector fabrication facility represents a transformative advancement in the production of ultra-high-purity Ge crystals and the development of next-generation detector technologies. By moving these processes underground, the facility effectively mitigates cosmogenic isotope production, a significant source of background noise generated during the manufacturing of Ge detectors at the Earth's surface. Reducing cosmogenic long-lived isotopes is critical for ensuring the sensitivity and performance of detectors used in next-generation Ge-based DM and \(0\nu\beta\beta\) decay experiments. The integration of crystal growth, mechanical processing, and detector fabrication within a single underground facility provides a controlled environment that supports the precise and high-quality production of detector-grade crystals. As the demand for more sensitive and reliable detectors continues to grow, this underground facility will be pivotal in meeting the stringent requirements of future scientific research, establishing itself as a key asset in the global effort to unlock the mysteries of the universe.

 \section{Acknowledgement} 
  This work was supported in part by NSF OISE 1743790, NSF PHYS 2310027, DOE DE-SC0024519, DE-SC0004768,  and a research center supported by the State of South Dakota.

\end{document}